\documentstyle[aps,manuscript]{revtex}
\begin{document}

\title{Soft spin waves in the low temperature thermodynamics of 
Pr$_{0.7}$Ca$_{0.3}$MnO$_{3}$}

\author{M. Roy$^{1}$, J. F. Mitchell$^{2}$, A. P. Ramirez$^{3}$, 
P. Schiffer$^{1}$\renewcommand{\thefootnote}{\alph{footnote}}
\footnote{schiffer.1@nd.edu}}

\address{$^{1}$Department of Physics, University of Notre Dame, Notre Dame, IN 46556}
 
\address{$^{2}$Material Science Division, Argonne National Laboratory,
 Argonne, IL 60439} 

\address{$^{3}$Bell Laboratories, Lucent Technologies, Murray Hill, NJ 07974 }

\maketitle
\begin{abstract}

We present a detailed magnetothermal study of Pr$_{0.7}$Ca$_{0.3}$MnO$_{3}$, 
a perovskite manganite in which an insulator-metal transition 
can be driven by magnetic field, but also by pressure, 
visible light, x-rays, or high currents.  We find that the 
field-induced transition is associated with an
enormous release of energy which accounts for its strong irreversibility.
In the ferromagnetic metallic state, specific heat and magnetization
measurements indicate a much smaller spin wave stiffness than that
seen in any other manganite, which we attribute to spin waves
 among the ferromagnetically ordered Pr moments.  The coupling between the Pr and Mn spins may also provide a basis for understanding the low temperature phase diagram of this most unusual manganite.

\end {abstract}
\newpage

The rare earth perovskite manganites (R$_{1-x}$A$_{x}$MnO$_{3}$) are
associated with a wide variety of fascinating physics due to the strong
coupling between their electronic, magnetic and lattice degrees of freedom.
Phenomena observed in these materials include $``$colossal$"$
magnetoresistance, real-space charge ordering, electronic phase separation,
and a diverse variety of magnetoelectronic ground states \cite{apr:97}.  
Although this entire class of materials displays unusual
behavior, one material, Pr$_{0.7}$Ca$_{0.3}$MnO$_{3}$
(PCMO), has been shown to display a particularly rich set of phenomena.
Like several other manganites, the resistivity of PCMO is reduced
by orders of magnitude in a magnetic field due to an irreversible transition
from an insulating antiferromagnetic to a metallic ferromagnetic
state \cite{jira:80,yosh:95,anan:99,tomi:96,lees:96}. 
What sets PCMO apart from the other manganites is that the metastable
insulating state also can be driven metallic by the application 
of light \cite{miya:97,cox:98}, pressure \cite{mori:97}, 
x-rays \cite{kiry:97}, or a high current \cite {asam:97}. Despite the strong 
recent interest in this wide range of unique phenomena in PCMO, there 
has been no clear understanding of why this particular manganite 
is so different from the others.  
	
We have performed a detailed study of the low temperature magnetothermal
properties of PCMO.  We find that there is an enormous release of heat at
the field-induced insulator to metal transition at low temperatures,
which explains the complete irreversibility of the transition.   
In the ferromagnetic (FM) state at low temperatures, our specific heat and
magnetization measurements indicate a ferromagnetic spin wave
stiffness which is far below that seen in other conducting manganites.  The data
can be explained most easily as a result of ferromagnetism among the moments
associated with the Pr ions, suggesting that the Pr magnetism is
crucial to understanding the unusual low temperature properties of this
material.

We studied both a ceramic sample of PCMO, synthesized by a standard
solid state technique, and a single crystal grown in a floating zone mirror
furnace. Both samples were judged to be single phase from x-ray diffraction studies.  The cation concentration ratio, as measured by plasma atomic 
emission spectroscopy, was also consistent with the nominal concentration. 
While we show only results from the single crystal, data from
the two samples were qualitatively and quantitatively consistent in all of
the studied properties, and both had low temperature phase diagrams (including  a regime of field-induced ferromagnetic metallic phase) consistent with previously published studies of PCMO \cite{tomi:96}.
 Magnetization was measured in a Quantum Design
SQUID magnetometer, and specific heat was measured by a semiadiabatic heat
pulse technique calibrated against a copper standard.  Magnetocaloric
measurements were made by temperature-controlling the
calorimeter at a few degrees above the surrounding
cryostat temperature while the field was swept.  We then measured 
the input power required to maintain constant temperature 
during the field sweep and attributed changes in the input 
power to magnetocaloric effects.

The field-temperature phase diagram of PCMO is shown in the inset of
figure~\ref{f1} based on earlier measurements \cite{tomi:96}.  
Upon cooling in zero 
field, PCMO first undergoes a charge-ordering transition at 
T$_{co}$ $\sim$ 230 K. Upon further cooling, the Mn spins order 
first into a pseudo-CE type antiferromagnetic state at T$_{AF}$ $\sim$ 150 K, 
and then into a canted antiferromagnetic (CAFM) state at 
a lower temperature T$_{CAFM}$ \cite {yosh:95}. This low
field phase is quite complex due to the incommensurability of the charge
order with the 30$\%$ Ca doping, and the resultant disorder leads to both
spin-glass-like behavior \cite{yosh:95,tomi:96}  and electronic phase
separation \cite{cox:98}.  Application of a sufficiently high magnetic
field at low temperatures induces a first order transition from the CAFM
phase to a conducting FM state, and there is a strong
hysteresis associated with this transition (the hatched
region in the inset indicates where the state is history-dependent).
For T $\lesssim$ 60 K, the transition is completely irreversible so that if the
material undergoes the field-induced FM transition it remains in a 
FM conducting state until the temperature is raised above 60 K.  In
figure 1 we demonstrate this irreversibility by plotting the 
magnetization vs. field after zero-field-cooling to T = 10 K and then 
sweeping the field up and down 0 $\rightarrow$ 7 T $\rightarrow$ -7 T
$\rightarrow$ 7 T. The reduced moment during the initial sweep up in 
field for H $<$ 4 T corresponds to the magnetization of the canted
antiferromagnetic phase, and the rise in magnetization during that 
sweep at $\sim$ 4 T corresponds to the phase transition. As seen in 
the figure, the magnetization maintains its
ferromagnetic nature during all subsequent field sweeps.

Since the field-induced phase transition is highly irreversible and
therefore first order, one expects an associated release of heat which should be observed in our magnetocaloric
measurements.  In figure 2 we show the raw data of a magnetocaloric
measurement, plotting the power input from the temperature controller (P)
vs. magnetic field after zero-field-cooling as the field is changed from 
0 $\rightarrow$ 9 T, 9 T $\rightarrow$ -9 T, and -9 T
$\rightarrow$ 9 T.  For H $\lesssim$ 1 T we see small rises 
and falls in P(H) during the initial sweep up in field. We attribute 
these reproducible features to heat release associated with the
spin-glass-like character of the zero-field-cooled CAFM state 
\cite{yosh:95,anan:99}, i.e. irreversible relaxation of the 
spin configuration during the initial field sweep \cite{mydo:93,tsui:99}
(a more detailed treatment of this behavior will be included in a future paper 
\cite{roy1:99}). The most prominent feature in the magnetocaloric data, 
however, is the large negative peak in P(H) during the initial
field sweep.  This peak corresponds to a reduction in the heat
required from the temperature controller to maintain constant temperature, 
and we attribute this reduction to the heat released at 
the CAFM-FM transition.  Note that there is no
similar peak observed in subsequent sweeps of the magnetic field, which is
consistent with the irreversibility of the transition.

The integrated magnitude of the peak in P(H), $Q = \int({P/(\frac{dH}{dt}))dH}$, is extraordinarily large at low temperatures - about 15 $\pm$ 1 J/mole. Within our resolution, $Q$ is independent of both the sweep rate between 6 G/sec and 24 G/sec and the
temperature for T $\lesssim$ 40 K, but Q gradually decreases for higher 
temperatures \cite{roy1:99}. Comparing the magnitude of Q to $\int{C(T)dT}$,
one finds that the heat release at the transition is sufficient to raise the
temperature of a perfectly thermally isolated sample from 5 to 17.5 K,
i.e. by more than a factor of 3. Indeed, as can be seen in the inset to
figure 2, when we simply sweep the field on our 0.22 gram sample on
the calorimeter, the temperature of the calorimeter
rises from 5 to 11 K.  The magnitude of Q can be understood
as a release of the  Zeeman energy associated with the CAFM  
phase relative to that in the FM phase.  This can be calculated from our 
measured value of M(H) as $\int{HdM}$ over the field range of the 
transition which is 10 $\pm$ 1 J/mole for T $\lesssim$ 40 K, in 
reasonable agreement with the measured values of Q  \cite{roy1:99}. 
The large magnitude of Q relative to the 
specific heat explains the complete irreversibility of the transition
at low temperatures, since there is no corresponding energetic advantage to
the CAFM phase at low fields which would drive the reversal of the transition.

We measured the specific heat as a function of temperature, C(T), at 0,
3, 6, and 9 tesla after cooling at those fields and at 0 and 3 tesla in the
FM state (induced by raising the field  temporarily to 9 T after
reaching low temperatures). As shown in figure 3 we find that we can fit 
the temperature dependence in the FM state with the simple form 
C(T) = $\beta$ T$^{3}$ + $\delta$ T$^{3/2}$ where the two terms 
correspond to the lattice specific heat and the specific heat associated 
with ferromagnetic spin waves respectively.  
While we cannot rule out the possibility of a
contribution to the specific heat which is linear in temperature (such as
would arise from either free electrons or a spin glass state), we find that
$C_{mag}$ = C(T) - $\beta$ T$^{3}$ has a power law temperature dependence 
with an  exponent of almost exactly 1.5 as shown in the log-log 
plot in the inset to figure 3. Our fitted values of $\beta$ vary with
field by only a few percent, while $\delta (H)$ varies by 65$\%$ implying 
that the field dependence of the specific heat originates almost 
entirely in the magnetic component. We directly measured the 
field-dependence of the specific heat, C(H) by zero-field cooling 
the sample and then measuring the specific heat every 0.25 T 
while sweeping the field from 0 $\rightarrow$ 9 T $\rightarrow$ 
-9 T $\rightarrow$ 9 T as shown in figure 4. 
During the initial sweep up in field, there is a sharp drop in C(H) 
at about 4 T corresponding to the CAFM $\rightarrow$ FM phase 
transition. On all subsequent sweeps, C(H) is quite 
smooth and monotonically decreases as the magnitude of the applied 
field increases. Since the fits to C(T) indicate that the lattice contribution 
to the specific heat ($\beta T^{3}$) is almost field-independent, we fit C(H) in
the FM state to a Heisenberg spin wave model as shown by the solid line 
in figure 4 \cite{kitt:64}:  

\begin{equation}
{C(H) = A+\frac{V_{mole}k_{B}^{5/2}T^{3/2}}{4\pi^{2}D^{3/2}}
\int\limits_{g\mu_{B}H/k_{B}T}^{\infty}
\frac{x^2 e^{x}}{{({e^{x}-1})}^2}{\sqrt{x-\frac{g\mu_{B}H}{k_{B}T}}}dx} 
\label{e1}
\end{equation}
where $V_{mole}$ is the molar volume and the only two fitting constants A and D are an offset to account for 
the lattice contributions and the stiffness constant of the spin wave
spectrum respectively \cite{dom}. While such a form can also fit C(H) in 
other ferromagnetic manganites, the magnitude of the field-induced suppression 
of C(H) is 15 times larger in PCMO than in the other ferromagnetic manganites
such as La$_{0.7}$Sr$_{0.3}$MnO$_{3}$ (shown as dashed line in figure 4) and
La$_{0.7}$Ca$_{0.3}$MnO$_{3}$ \cite{roy:99}. This larger suppression of 
C in a field implies that the magnetic specific heat is much larger in PCMO
than in the other materials and consequently that the spin wave spectrum
is much softer. The fit to C(H) of PCMO
yields a value of D = 28.0 $\pm$ 0.3 meV-\AA$^{2}$.  This is a
factor of 5-6 smaller than that obtained from neutron scattering studies of
other ferromagnetic metallic manganites \cite{okuda:00} including La$_{0.70}$Sr$_{0.30}$MnO$_{3}$ \cite{mart:96}, 
La$_{0.67}$Ca$_{0.33}$MnO$_{3}$ \cite{lynn:96}, Pr$_{0.63}$Sr$_{0.37}$MnO$_{3}$
\cite{fern:98}, and La$_{0.70}$Pb$_{0.30}$MnO$_{3}$ \cite{perr:96} the 
first two of which are in good agreement with the measured 
field-dependence of C(H) \cite {roy:99}.

The low value of D from the fits to C(H) can be tested in two
ways. First, the parameter $\delta$ obtained from our fit to C(T) at H = 0 yields an 
estimate of D, since $\delta = 0.113 k_{B} V_{mole} {\left(\frac{k_{B}}{D}
\right)}^{3/2}$ within the Heisenberg spin wave model \cite{kitt:64}.  Our value of $\delta$ corresponds to D = 25.3 $\pm$ 0.5
meV-\AA$^{2}$, in reasonable agreement with the fit to C(H). Second, the 
very soft spin wave stiffness in the ferromagnetic state can also be
tested by fitting our magnetization data, since the magnetization per unit volume at low temperatures 
within the Heisenberg model is given by \cite{roy:99,kunz:60,hend:69,smol:97}, 

\begin{equation}
{M(0,H)-M(T,H)={g\mu_{B}}\left(\frac{k_{B}T}{4 \pi D}\right)^{3/2} f_{3/2}(g \mu _{B}(H-NM)/k_{B}T)} 
\label{e2}
\end{equation}
where ${f_{p}(y) = \sum_{n=1}^{\infty} \frac{e^{-ny}}{n^{p}}}$, and
NM is the demagnetization field and M is the magnetization. 
Such a fit to M(T) (shown in Figure 5 as the solid 
line) reproduces our data with the exception of the very 
lowest temperature points which are suppressed by $\sim$ 0.3$\%$ 
possibly due to a minute presence of a second phase (this feature 
is also seen in the polycrystalline sample). A fit to the data 
using equation \ref{e2} yields D = 25.1$ \pm$ 2.1 meV-\AA$^{2}$ 
which is in agreement with our other two values.

We now discuss a possible origin of the very low ferromagnetic
spin wave stiffness constant in PCMO. There is no reason to expect 
that the Mn-Mn ferromagnetic double exchange interaction should be 
a factor of 5 weaker in this compound than
in all of the other ferromagnetic metallic manganites (in fact studies
have shown that D at low temperatures is remarkably independent of 
ordering temperature in the manganites \cite{lynn:96}), and recent neutron scattering
measurements of D for the Mn spins in PCMO yield D $\sim$ 150 meV-\AA$^{2}$
\cite{fern:99}.  While one might invoke electronic phase separation to explain the
 thermodynamic data, it is difficult
to imagine that phase separation could reproduce quantitative 
agreement with the Heisenberg model consistently for the three different data sets.  
Furthermore, increasing the field should increase the proportion of the ferromagnetic phase
which already includes $\sim$ 90 $\%$ of the sample based on the magnetization data.  Thus, 
for $C_{mag}$ to decrease as strongly with field as we observe, the residual phase would
need to have an associated specific heat about 100 times larger than is typical for 
antiferromagnetic manganites \cite {roy:99}.  A more plausible explanation is suggested
by the data of Cox $\it{et}$ $\it{al.}$ who 
reported that there is a ferromagnetic moment associated with the Pr ions for 
temperatures below T$_{c}$ = 60 K \cite{cox:98}. We propose 
that the spin waves on this ferromagnetic sublattice are responsible
for our data, and we suggest that the Pr magnetism could be critical to the low temperature 
thermodynamics of PCMO. Softer spin waves associated
with the Pr moment would have a much larger associated specific 
heat than the Mn spins and therefore would dominate both C$_{mag}$(T) and
C(H). A careful examination of M(T) for our samples at low fields 
shows a distinct rise at T $\sim$ 50-60 K (Figure 5 inset), and C(T) also shows a peak 
there, confirming the existence of the ordering transition 
\cite{roy1:99}. Furthermore, in all studies of the 
magnetization of PCMO, the saturation moment at high fields and 
low  temperature \cite{lees:96,tomi:95,thom:99} is
at least 10$\%$ higher than that of other ferromagnets, such as 
La$_{0.70}$Sr$_{0.30}$MnO$_{3}$ and La$_{0.70}$Ca$_{0.30}$MnO$_{3}$
\cite{smol:97,thom:99}.
An additional ferromagnetic moment such as that from the Pr is required to
explain this excess magnetization since the Mn spins in PCMO display 
considerable canting even at high field \cite{yosh:95}.

The ferromagnetism among the Pr ions also appears to affect the remarkable
phase behavior of this material, since the ferromagnetic
T$_{c}$ $\sim$ 60 K  for the Pr corresponds to the temperature
below which the field-induced and pressure-induced CAFM-FM transitions 
become irreversible.  Furthermore, this is also the temperature above 
which the x-ray induced metallicity is
quenched \cite{kiry:97}.  It thus appears that the Pr ferromagnetism destabilizes 
the CAFM phase relative to the FM phase, so that in zero
field the CAFM phase is stable for T $>$ 60 K but then becomes unstable to the FM phase
below T $\sim$ 60 K when the Pr spins order.  The free energy difference between the CAFM 
and FM state
in zero field could be small enough that the FM state does not nucleate spontaneously at
such low temperatures, but an applied magnetic field would make the FM state 
energetically more favorable and induce the first order phase transition.
In this scenario, both the Pr ferromagnetism and its coupling to the Mn moments
 are crucial to understanding the physics 
and are therefore inseparable from the numerous unique phenomena 
observed in PCMO \cite{miya:97,cox:98,mori:97,kiry:97,asam:97}. The 
importance of rare-earth magnetism to the explanation of the unusual
properties of the manganites has been largely ignored in previous studies, 
and at least in the case of PCMO it appears that this
assumption is not justified.

We gratefully acknowledge helpful discussions with P. Dai and J. A. Fernandez-Baca, and financial support from NSF grant DMR 97-01548,
the Alfred P. Sloan Foundation and the Dept. of Energy, Basic Energy
Sciences-Materials Sciences under contract \#W-31-109-ENG-38.

\begin{figure}
\caption{Zero-field cooled magnetization of PCMO as a function of
field at T $=$ 10 K. The solid line is the initial sweep up in field, the open
circles and the dotted line show the subsequent field sweeps down and up
respectively. The inset shows the low temperature field-temperature 
phase diagram~\protect\cite{tomi:96} where the shaded region indicates 
the history 
dependent region.}  
\label{f1}
\end{figure}

\begin{figure}
\caption{Magnetocaloric measurement of PCMO, showing the power,
P(H), required to maintain the temperature of the calorimeter at T = 9.000
$\pm$ 0.025  while the surrounding cryostat was at 7.5 K.  The data were taken
after the sample was zero-field-cooled and the field was changed from 
0 $\rightarrow$ 9 T (solid line), 9 T $\rightarrow$ -9 T (dashed line),  
and -9 T $\rightarrow$ 9 T (dotted line) at the rate of 6 gauss/sec (only
H $>$ 0 data are shown). The inset illustrates the temperature rise 
of the calorimeter upon the initial sweep up in the field when it is 
not temperature controlled.}
\label{f4}
\end{figure}

\begin{figure}
\caption{Specific heat as a function of temperature in the FM state 
as described in the text, H = 0 (open circle), H = 3 T (open down-triangle), 
H = 6 T (open up-triangle) and  H = 9 T (open square). The solid lines are fits 
as discussed in the text. The inset shows the magnetic specific heat of the FM 
state at H = 0 and 9 T on a log-log scale. The solid lines have slope 
of 1.5, the dashed and the dotted lines have slopes of 2 and 1 respectively.} 
\label{f3}
\end{figure}

\begin{figure}
\caption{Zero field-cooled C(H), when the
field is initially swept from 0 $\rightarrow$ 9 T (closed circles), 
9 T $\rightarrow$ -9 T (open circles) and -9 T $\rightarrow$ 9 T 
(open triangles) at T = 5.5 K. The solid lines are the fits to 
the data as discussed in the text. The dotted line shows C(H) for La$_{0.7}$Sr$_{0.3}$MnO$_{3}$ with a constant offset to match
C(H = 0) for PCMO.}
\label{f2}
\end{figure}

\begin{figure}
\caption{The low temperature magnetization of PCMO as a function
of temperature at H = 7 T. The left inset illustrates the weighted fit for
the calculation of D as discussed in the text.  The right inset illustrates 
an additional feature in M(T) at low temperatures ( T $<$ 60 K) 
presumably associated with Pr ordering, where the straight line is a guide 
to the eye.}
\label{f5}
\end{figure}


\begin{references}
\bibitem{dom} We fit the data for $|$H$|$ $>$ 1 T so as to avoid domain effects which are evident in both the magnetization and magnetocaloric data.
\end{references}
\end {document}